\documentclass[
  aps,
  prb,
  twocolumn,
  superscriptaddress
]{revtex4-2}
\usepackage{amsmath,amssymb,bm}
\usepackage{graphicx}
\usepackage{dcolumn}
\usepackage{color}
\usepackage{accents}
\usepackage{mathtools}
\usepackage{ulem}
\usepackage{url}
\usepackage{hyperref}

\begin{document}

\title{Reentrant Superconductivity in Zeeman Fields}
\date{\today}

\author{Tomoya Sano}
\thanks{These authors contributed equally to this work.}
\affiliation{Department of Applied Physics, Hokkaido University, Sapporo 060-8628, Japan}
\author{Kota Tabata}
\thanks{These authors contributed equally to this work.}
\affiliation{Department of Applied Physics, Hokkaido University, Sapporo 060-8628, Japan}
\author{Satoshi Ikegaya}
\affiliation{Department of Applied Physics, Hokkaido University, Sapporo 060-8628, Japan}
\author{Yasuhiro Asano}
\affiliation{Department of Applied Physics, Hokkaido University, Sapporo 060-8628, Japan}

\begin{abstract}
We propose a theoretical model for a superconductor that exhibits 
the reentrant superconductivity in Zeeman fields. 
The Bogoliubov-de Gennes Hamiltonian includes three vectors 
in spin space: a $d$ vector of a spin-triplet superconducting state, 
a potential representing spin-orbit interactions, and a Zeeman field. 
When the three vectors are perpendicular to one another, the 
spin-orbit interaction suppress superconductivity in weak Zeeman fields 
and enhances superconductivity in strong Zeeman fields.
The instability (stability) of superconducting state is characterized by the appearance of
odd-frequency (even-frequency) Cooper pairs.
\end{abstract}

\maketitle


\textit{Introduction}: An external magnetic field suppresses superconductivity in 
two ways: orbital effect and spin Zeeman effect.
The orbital effect common to all superconductors (SCs) fluctuates 
the phase of the superconducting condensate and increases the free-energy.
In layered SCs and monolayer SCs under a magnetic field applied
parallel to the layers, the orbital effect is negligible and 
only the Zeeman effect modifies the superconducting states.
A Zeeman field always suppresses spin-singlet superconductivity.
The critical magnetic field $H_c$ at zero temperature is called  
Pauli limit $H_{\mathrm{P}}$.~\cite{chandrasekhar:apl1962,clogston:prl1962}.
For spin-triplet superconductivity, 
a Zeeman filed $\bm{H}$ parallel to a $\bm{d}$ vector suppresses superconductivity, whereas
that perpendicular to the $\bm{d}$ does not affect the superconducting state.
In experiments, Zeeman-field-induced superconductivity (ZFIS) or reentrant superconductivity
have been observed various materials such as an organic SC 
$\lambda-\mathrm{(BETS)}_2\mathrm{FeCl}_4$~\cite{uji:nature2001},
Europium compounds~\cite{meul:prl1984}, UCoGe\cite{aoki:jpsj2009}, LaTiO$_3$-KTaO$_3$ 
interface~\cite{maryenko:arxiv2025}, and 
UTe$_2$ \cite{ran:natphys2019,aoki_ute2:jpsj2020,helm:natphys2024}. 
The last one has attracted attention these days as a promising candidate of a spin-triplet SC.
However, our understanding of ZFIS remains extremely limited.
Only the Jaccarino-Peter compensation~\cite{jaccarino:prl1962} is the widely accepted 
as the mechanism explaining the transition to a superconducting phase in high Zeeman fields. 
Thus, another scenario for explaining the ZFIS is desired to understand physics 
behind the experimental findings.
We will address such an issue in this paper.

To demonstrate the ZFIS, we introduce the third vector in spin space $\bm{\alpha}$
that describes spin-orbit interactions (SOIs). 
It has been established that SOIs increase $H_c$ of spin-singlet SCs beyond the Pauli limit
\cite{klemm:prb1975,tedrow:prb1982}. 
A recent paper reported very complicated phase diagrams for spin-triplet SCs 
with SOIs~\cite{ma:arxiv2025}.
A Zeeman field and a SOI induce additional pairing correlations. Although they 
\textsl{do not} form the pair potential, they govern the stability of superconductivity 
depending on their frequency symmetry classes.
Odd-frequency Cooper pairs decrease the transition 
temperature $T_c$~\cite{asano:prb2015,ramires:prb2018,triola:annphys2020}
because they decrease the superfluid weight.
On the other hand, even-frequency Cooper pairs increase $T_c$.~\cite{sano:prb2026}
In this paper, we show that spin-singlet Cooper pairs whose correlation function is proportional to 
a scalar chiral product $\bm{\alpha}\times\bm{V} \cdot \bm{d}$ assist spin-triplet 
superconductivity in sufficiently strong Zeeman fields.

\begin{figure*}[thb]
  \centering
  \includegraphics[width = 1.6\columnwidth]{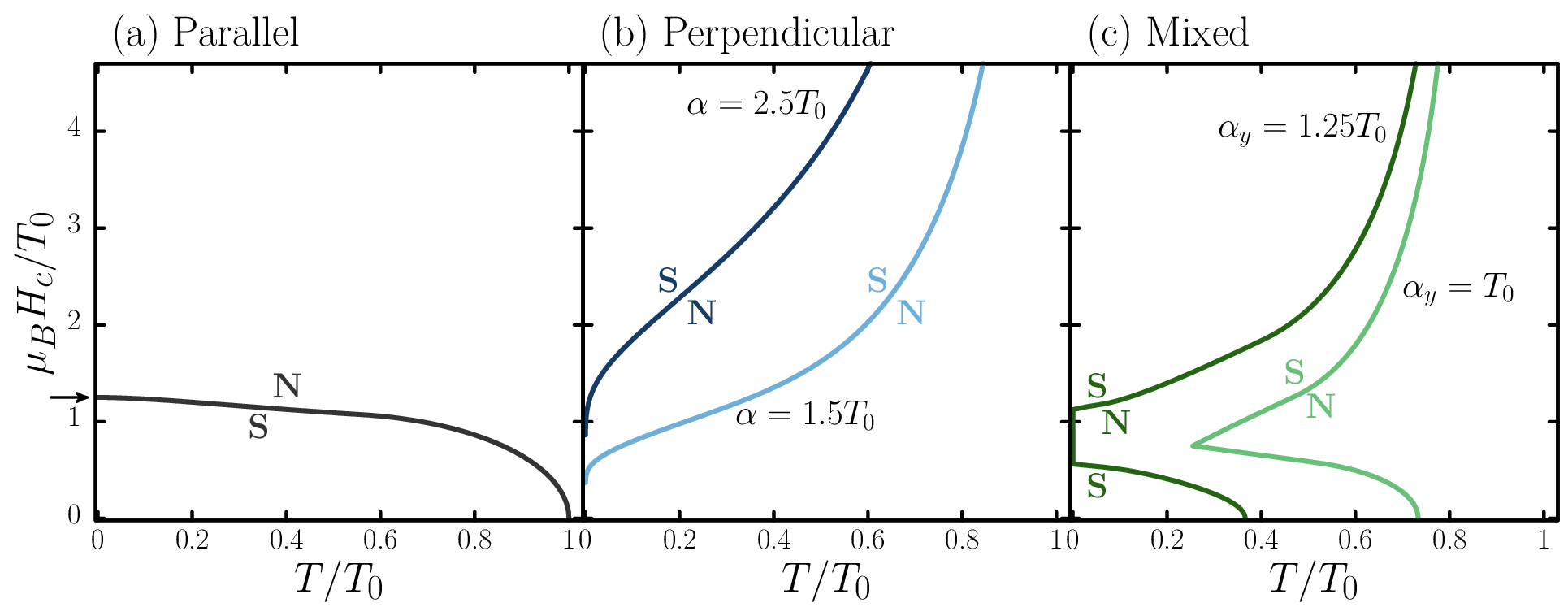}
  \caption{
    The critical magnetic field $H_{c}$ is plotted as a function of temperature $T$.
(a) All the three vectors are aligned in parallel, $\bm{d} \parallel \bm{\alpha} \parallel \bm{H}$.
The arrow at the vertical axis indicates the Pauli limit $H_{\mathrm{P}}$.
(b) The three vectors are perpendicular to one another, $\bm{d} \perp \bm{\alpha}$, 
$\bm{d}\perp \bm{H}$, and $\bm{\alpha} \perp \bm{H}$.
At $H=0$, superconductivity is absent because of $\alpha > \mu_{\mathrm{B}} H_{\mathrm{P}}$.
(c) The mixed states with $\alpha_x=0.75T_0$ exhibit the reentrant superconductivity. 
The two superconducting phases tend to separate from each other for large $\alpha_y$.   
  } \label{fig1}
\end{figure*}

\textit{Model}: We consider an electronic structure in two dimension, 
where two Fermi surfaces surround at certain points $\bm{K}$ and $-\bm{K}$ in momentum space.  
We describe such two Fermi surfaces in terms of two valleys. 
The normal state Hamiltonian is given by
\begin{align}
  \check{H}_{\mathrm{N}} &(\bm{k}) \label{eq:normal state hamiltonian}
  =
  \hat{\xi}_{\bm{k}} \hat{\sigma}_{0}
  +
  \bm{\alpha} \cdot \hat{\bm{\sigma}} \hat{\rho}_{z}
  +
  \bm{V} \cdot \hat{\bm{\sigma}} \hat{\rho}_{0} , \\
  \hat{\xi}_{\bm{k}} \label{eq:xipm}
  &=
  \frac{1}{2 m} \left( \bm{k} \hat{\rho}_{0} - \bm{K} \hat{\rho}_{z} \right)^{2} - \mu,  
  \quad \bm{V}= \mu_{\mathrm{B}} \bm{H}
\end{align}
where $\mu$ is the chemical potential and 
$\bm{H}$ represents a Zeeman field with $\mu_{\mathrm{B}}$ being the Bohr magneton.
We use the unit of $k_{B} = \hbar = c = 1$, where $k_{B}$ is the Boltzmann constant and 
$c$ is the speed of light.
The Pauli matrices in spin and valley spaces are denoted by 
$\hat{\bm{\sigma}} = (\hat{\sigma}_{x}, \hat{\sigma}_{y}, \hat{\sigma}_{z})$ and 
$\hat{\bm{\rho}} = (\hat{\rho}_{x}, \hat{\rho}_{y}, \hat{\rho}_{z})$, respectively.
The unit matrices in the two spaces are denoted by $\hat{\sigma}_{0}$ and $\hat{\rho}_{0}$.
We assume that the SOI $\bm{\alpha}$ is independent of $\bm{k}$ and changes its sign in the two valleys. 
The time-reversal operation is given 
by $\mathcal{T} = i \hat{\sigma}_{y} \hat{\rho}_{x} \mathcal{K}$, where $\mathcal{K}$ means the complex conjugation plus $\bm{k} \to - \bm{k}$.
The particle-hole conjugation is represented by 
$\undertilde{\check{H}}_{\mathrm{N}} (\bm{k}) = \check{H}^{*}_{\mathrm{N}} (- \bm{k})$ as usual.
Two electrons at the different valleys form a spin-triplet Cooper pair.
The pair potential for such a pair is represented as
\begin{align}
  \check{\Delta} = i \, \bm{d} \cdot \hat{\bm{\sigma}} \, \hat{\sigma}_{y} \,  i \hat{\rho}_{y},
  \label{eq:delta}
\end{align}
where $\bm{d}$ is independent of $\bm{k}$. Instead of $\bm{d}$ being odd-parity function,  
making valley-parity odd preserves the antisymmetric property of the pair potential
derived from the Fermi-Dirac statistics of electrons.
We note that all Cooper pairs in this model belong to even-parity $s$-wave symmetry class.
Although the similar electronic structures are realized in 
transition-metal dicalcogenides~\cite{saito:natphs2016,xi:natphs2016,barrera:natcom2018}, 
we do not focus on any specific materials in this paper.  
The advantage of the model is briefly summarized as follows.
In single-band SCs, $\bm{\alpha}$ and $\bm{d}$ are an odd-parity function 
depends on $\bm{k}$. In our model, they are described by the $\bm{k}$ independent 
potentials which change the sign under interchanging the valley indices. 
As a result, magnetic properties of a SC is purely derived from the relative 
vector configuration among $\bm{d}$, $\bm{\alpha}$, and $\bm{H}$. 
At the end of this paper, we will confirm that main conclusions of this paper
are valid also in single-band odd-parity spin-triplet SCs.

We solve the Gor'kov equation for the Bogoliubov-de Gennes (BdG) Hamiltonian
\begin{align}
  &\left[ i \omega_{n} - H_{\mathrm{BdG}} (\bm{k}) \right]
  \begin{bmatrix}
    \check{{G}} & 
    \check{{F}} \\
    - \undertilde{\check{{F}}} & 
    - \undertilde{\check{{G}}} 
  \end{bmatrix}_{(\bm{k}, \omega_{n})}
  = 1 ,\label{eq:bdg} \\ 
  &H_{\mathrm{BdG}} (\bm{k})
  =
  \begin{bmatrix}
    \check{H}_{\mathrm{N}} (\bm{k}) & \check{\Delta} \\
    - \undertilde{\check{\Delta}} & - \undertilde{\check{H}}_{\mathrm{N}} (\bm{k})
  \end{bmatrix} ,
\end{align}
where $\omega_{n} = (2n + 1) \pi T$ is a Matsubara frequency with $T$ being a temperature.
The analytical expression of the anomalous Green's function is supplied in Eq.~(S4)
in Supplemental Material~\cite{SM}.
By substituting the anomalous Green's function into the gap equation in Eq.~(S10), 
the pair potential is calculated in a self-consistent way.

The thermodynamics of a superconductor near the transition temperature is described well by the 
Ginzburg-Landau free-energy,
\begin{align}
\Omega_S
=&- \frac{1}{2} T  \frac{1}{V_{\mathrm{vol}}} \sum_{\bm{k}}  \sum_{j=1}^4
\log \left( 2 \cosh \left[ \frac{ E_{\bm{k},j}}{2T} \right] \right) 
+ \frac{\bm{d}^2}{g}, \label{eq:fe1} \\
\approx &  \, a\, d^2 + b\, d^4 + \mathrm{h.o.t}, \label{eq:gl}
\end{align}
where $E_{\bm{k},j}$ is the eigen energy of the BdG Hamiltonian and $d$ 
in Eq.~\eqref{eq:gl} is the amplitude of 
the $\bm{d}$ vector.
The coefficient $a$ is proportional to the linearized gap equation 
 which is 0 at $T = T_c$ and negative for $T < T_c$ as shown below 
in Eqs.~\eqref{eq:para_gap} and \eqref{eq:perp_gap}. 
Magnetically active potentials generate the pairing correlations belonging to 
various symmetry classes as shown in Eq.~(S4). Among them,   
only the spin-triplet odd-valley parity pairing correlation form the pair potential 
through the gap equation. 
Hereafter, we refer to such pairing correlation as principal correlation.  
Although the decrease in $T_c$ certainly provides a quantitative measure of the instability of the superconducting state, the information obtained from the gap equation is limited to the amplitude of the principal correlation.
To compensate for the lack of information, we discuss how 
remaining correlations in Eq.~(S4) stabilize or destabilize the superconducting 
states by using the superfluid weight defined by 
\begin{align}
Q_F= T\sum_{\omega_n} \int d \xi_{\bm{k}}\, \frac{1}{2}\mathrm{Tr}[-F \undertilde{F}]_{(\bm{k}, \omega_{n})}.
\end{align}
In SM~\cite{SM}, a relation of $Q_F$ to the Meissner screening length is also explained.
We mainly discuss the superfluid weight at the lowest order of $\bm{d}$,
\begin{align}
q(T,H) = \left. \frac{Q_F(T,H)}{d^2} \right|_{d \to 0}. \label{eq:smallq_def}
\end{align}
Here we summarize several general features of $q$. 
The superfluid weight of induced odd-frequency pairing correlations is negative.
As a consequence, the appearance of odd-frequency pairs makes the superconducting states 
unstable and decreases $T_c$~\cite{asano:prb2015}. 
Such situation is indirectly reflected in the coefficient $a$ in Eq.~\eqref{eq:gl}. 
The coefficient $b$ in Eq.~\eqref{eq:gl} is proportional to $q(T_c, H_c)$~\cite{sato:prb2024}.
Therefore, the transition to the superconducting phase becomes a first-order 
for $q(T_c, H_c)<0$.

\begin{table}[ttt]
\caption{
Symmetry classification of the pair correlation functions.
The top row corresponds to the spin-triplet pairing correlation that is linked to the pair potential 
through the gap equation and is referred to as principal pairing correlation.
It appears at the first term in Eqs.~\eqref{eq:para_f2} and \eqref{eq:perp_f2}.
The second and third low represent 
 odd-frequency pairing correlations appearing at the second term 
 in Eqs.~\eqref{eq:para_f2} and \eqref{eq:perp_f2}.
They make the superconducting unstable.
The cooperative effect between a Zeeman filed and a SOI induces
the spin-singlet pairing correlation at the bottom row.
All the Cooper pairs belong to even-momentum-parity $s$-wave symmetry class.
}
\begin{ruledtabular}
\begin{tabular}{clcl}
 frequency & spin ($\times i \hat{\sigma}_y$)  & valley-parity & \\
\colrule
 even & triplet  $\bm{d} \cdot \hat{\bm{\sigma}}$ & odd $\hat{\rho}_y$ & principal \\
 odd & singlet   $\bm{d} \cdot \bm{H} $  & odd $\hat{\rho}_y$ & induced $F_\parallel$ \\
 odd & triplet $\bm{\alpha} \times \bm{d} \cdot \hat{\bm{\sigma}}$ & even  $\hat{\rho}_x$ & induced $F_\perp$\\
 even & singlet $\bm{\alpha} \times \bm{d}  \cdot \bm{H}$   & even $\hat{\rho}_x$ & induced $F_\perp$\\
\end{tabular}
\end{ruledtabular}
\label{table1}
\end{table}

\begin{figure}[thb]
  \centering
  \includegraphics[width = \columnwidth]{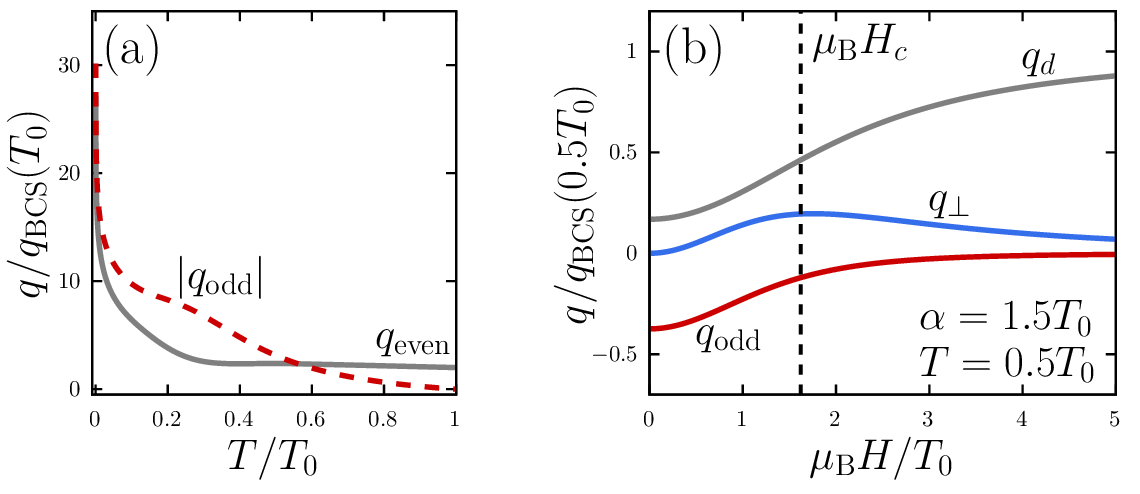}
  \caption{
 In (a), the two superfluid weights for the parallel configuration 
are plotted along the phase boundary in Fig.~\ref{fig1}(a).
In (b), the three superfluid weights are plotted as a function of Zeeman fields 
for the perpendicular configuration, where the vertical doted line indicates the critical field
at $T=0.5T_0$ and $\alpha=1.5T_0$ in Fig.~\ref{fig1}(b). 
  } \label{fig:q}
\end{figure}

\textit{Results}: 
At first, we discuss a configuration $\bm{\alpha} \parallel \bm{H} \parallel \bm{d}$, 
where the three vectors align in the same direction.
The anomalous Green's function is given in Eq.~(S13) in SM.
To solve the linearized gap equation, the summation over $\bm{k}$ is necessary. 
The shift of the wavenumber $\bm{k} \to \bm{k} \mp \bm{K}$ for $\xi_\pm$ 
remains the results unchanged because $\bm{\alpha}$ and $\bm{d}$ 
are independent of $\bm{k}$. 
By changing the summation over $\bm{k}$ 
to the integration over $\xi_{\bm{k}} \mp \alpha \hat{\rho}_z$, we reach an expression  
of
\begin{align}
  \frac{1}{V_{\mathrm{vol}}} &\sum_{\bm{k}} 
  \check{F}_\parallel (\bm{k}, \omega_{n}) \nonumber\\ 
  =&\frac{N_0 \pi}{|\omega_n|} 
  \left[ \frac{\omega_n^2}{\omega_n^2 +  V^2} \bm{d} \cdot \hat{\bm{\sigma}} 
- \frac{i \,\omega_n\, \bm{d} \cdot \bm{V} }{\omega_n^2 + V^2}   
  \right] 
   \hat{\sigma}_{y} \hat{\rho}_y,\label{eq:para_f2}
\end{align}
where $N_0$ is the density of states at the Fermi level per spin. 
The SOI modifies the band-dispersion 
only slightly in this configuration. 
The first term corresponds to the principal correlation.
A Zeeman field generates the odd-frequency pairing correlation at the second term.
The two pairing correlations are listed at the first and second rows in Table~\ref{table1}.
We find that the structure of the anomalous Green's function 
in Eq.~\eqref{eq:para_f2} is essentially the same as that 
for a spin-singlet superconductor under a Zeeman field by 
changing $\bm{d} \cdot \hat{\bm{\sigma}} \to \Delta$ and $\bm{d} \cdot \bm{V} 
\to \Delta \bm{V} \cdot \hat{\bm{\sigma}}$.
In fact, the resulting linearized gap equations in the two cases 
\begin{align}
\ln\frac{T}{T_0}
+ 2\pi T \sum_{\omega_n>0} \frac{1}{\omega_n}\left[1-\frac{\omega_n^2}{\omega_n^2+ V^2}\right]=0,
\label{eq:para_gap} 
\end{align}
are identical to each other.
Here $T_0$ is the transition temperature at $\bm{H}=\bm{\alpha}=0$.
The phase boundary between the normal and superconducting states in Fig.~\ref{fig1}(a) is 
determined by two methods: solving the gap equation in Eq.~\eqref{eq:para_gap} and 
finding the minimum of the free-energy in Eq.~\eqref{eq:fe1}.
Two results are identical to each other when the transition to the superconducting states is a second-order.
For first-order transitions, the phase boundary is obtained from the free-energy minima.
The arrow on the vertical axis indicate the Pauli limit 
$\mu_{\mathrm{B}}\, H_{\mathrm{P}}$~\cite{chandrasekhar:apl1962,clogston:prl1962}.
The superfluid weight of the principal correlation $q_{d}(T, H_c)$ and that for the induced odd-frequency 
correlation $q_{\mathrm{odd}}(T, H_c)$ are plotted as a function of temperature in 
Fig.~\ref{fig:q}(a), 
where $H_c$ is obtained from the data points on the phase boundary in Fig.~\ref{fig1} (a) and
the vertical axis is normalized to $q_{\mathrm{BCS}}$ in Eq.~(S20) in SM. 
The transition to the superconducting state in Fig.~\ref{fig1}(a) becomes a first-order for
$T < 0.556 T_0$~\cite{maki_tsuneto:ptp1964,sarma:jpcs1963} because
odd-frequency pairing correlation decreases $q(T_c, H_c)$ to be negative 
as demonstrated in Fig.~\ref{fig:q}(a). 
A Zeeman field seriously suppresses spin-triplet superconductivity in the parallel configuration.

Secondly, we consider  
a configuration $\bm{\alpha} \perp  \bm{d} $, $\bm{\alpha} \perp \bm{H}$, 
and $\bm{H} \perp \bm{d}$, where the three vectors are perpendicular to one another.
The anomalous Green's function near the transition temperature is supplied in Eq.~(S15) in SM~\cite{SM}.
Here we show the results after carrying out the summation over $\bm{k}$ by shifting the 
wavenumber $\bm{k} \to \bm{k}\mp \bm{K}$ for $\xi_\pm$,
\begin{align}
  \frac{1}{V_{\mathrm{vol}}} \sum_{\bm{k}} 
  &\check{F}_\perp (\bm{k}, \omega_{n})
  =\frac{N_0 \pi }{|\omega_n|(\omega_n^2 + \alpha^2 + V^2)} \nonumber\\ 
&\times \left[
(\omega_n^2 + V^2) \bm{d} \cdot \hat{\bm{\sigma}} +
\omega_n ( \bm{\alpha} \times \bm{d}) \cdot \hat{\bm{\sigma}}\hat{\rho}_z  \right.\nonumber\\
& \left.+ i  ( \bm{\alpha} \times \bm{d}) \cdot \bm{V} \hat{\rho}_z
\right] 
\hat{\sigma}_y \, \hat{\rho}_y. \label{eq:perp_f2}
\end{align}
The first term is the principal pairing correlation.
The odd-frequency pairing correlation at the second term is induced by the SOI and
suppresses superconductivity~\cite{frigeri:njp2004}. 
We will show that the third term listed at the bottom row in Table~\ref{table1} plays a key role in the ZFIS. 
The linearized gap equation results in
\begin{align}
\ln\frac{T}{T_0}
&+ 2\pi T \sum_{\omega_n>0} \frac{1}{\omega_n}
\left[1-\frac{\omega_n^2+ V^2}{\omega_n^2+V^2+\alpha^2}\right]=0.\label{eq:perp_gap} 
\end{align}
 Equation~\eqref{eq:perp_gap} suggests that
 the large SOI $\alpha > \mu_{\mathrm{B}} H_{\mathrm{P}}$ deletes superconductivity at $\bm{H}=0$.
In the $H-T$ phase diagram in Fig.~\ref{fig1}(b), superconductivity is absent at $\bm{H}=0$ 
because we choose such large SOI as $\alpha=1.5T_0$ and $2.5T_0$.
Equation~\eqref{eq:perp_gap} also indicates that a Zeeman field screens such negative effects due to the SOI.
Fig.~\ref{fig1}(b) shows that the superconducting phase appears in large Zeeman fields and 
that the critical temperature increases with increasing Zeeman fields.
We explain the reasons below.
The analytic expressions of the superfluid weight of the principal correlation $q_d$, that of
odd-frequency pairs $q_{\mathrm{odd}}$, and that of spin-singlet pairs $q_{\perp}$ are supplied
in Eqs.~(S22)-(S23) in SM.
They are plotted as a function of the Zeeman potential in Fig.~\ref{fig:q}(b), where
we choose $\alpha=1.5 T_0$ and fix the temperature at $T=0.5T_0$. 
The results show that $q_\perp $ compensates the negative weight of $q_{\mathrm{odd}}$ 
in Zeeman fields. 
Namely, the induced spin-singlet Cooper pairs stabilize the spin-triplet 
superconductivity in Zeeman fields. 
The increase of $q_d$ with Zeeman fields is a result of such compensation, 
which explains the ZFIS described by the gap equation in Eq.~\eqref{eq:perp_gap}.
We conclude that the interplay between the SOI and the Zeeman field realizes ZFIS.
Eq.~\eqref{eq:perp_gap} suggests that $T_c$ goes to the $T_0$ for $V \gg \alpha $.
However, our phenomenological theory cannot predict such limiting behavior.
We have assumed in high magnetic fields that the attractive interaction between two electrons 
remains unchanged and that other conduction bands do not come to the Fermi level.

Finally, we consider a mixed situation between the parallel and the perpendicular configurations,
\begin{align}
\bm{\alpha} =& \alpha_x \bm{e}_x + \alpha_y \bm{e}_y,\quad
\bm{d} =d_x \bm{e}_x + d_z \bm{e}_z, \label{eq:mix} 
\end{align}
and $\bm{H}= H \bm{e}_x$, where both the SOI and the pair potential have two components.
Based on $H-T$ phase diagram in Figs.~\ref{fig1}(a) and (b) 
and the interpretation of the results 
by using odd-frequency Cooper pairing correlations, it is possible to predict
the reentrant superconductivity in the mixed configuration in Eq.~\eqref{eq:mix}.
At $H=0$, the odd-frequency pairing correlation 
proportional to $\bm{\alpha} \times \bm{d} \cdot \hat{\bm{\sigma}}$ make the superconducting 
states unstable. Both $\alpha_x$ and $\alpha_y$ suppress the pair potential $d_z$, 
whereas only $\alpha_y$ suppresses $d_x$.
As a result, single-component superconducting states with $\bm{d}=(d_x, 0,0)$ 
would be realized for small $H$.
In high Zeeman fields, the pair potential $d_x$ vanishes due to the odd-frequency 
correlation proportional to $\bm{d} \cdot \bm{H}$ and the superconductivity 
due to $d_z$ would be stabilized by the spin singlet pairing correlation proportional to
$\bm{\alpha}\times\bm{d} \cdot \bm{H}$. 
The phase diagram obtained from the minima of the free-energy shown 
in Fig.~\ref{fig1}(c) are consistent with the predictions, where we 
choose $\bm{g}=(g, 0, g)$ in Eq.~(S11) and fix $\alpha_x$ at 0.75 $T_0$. 
With increasing Zeeman fields from zero, the transition temperature 
first decreases for both $\alpha_y/T_0=1.0$ and 1.25.
The results for $\alpha_y=1.25T_0$ show that superconductivity vanishes 
for $ 0.55 T_0 < \mu_{\mathrm{B}}H < 1.12 T_0$ even at $T=0$.
Another single component superconductivity with $\bm{d}=(0,0,d_z)$ 
appears for high Zeeman fields $ 1.12 T_0 < \mu_{\mathrm{B}}H$. 
 For $\alpha_y=T_0$, 
the low-field phase and high-field phase are connected at the vending point.
 However, the $\bm{d}$ vector has only one component in the two phases: 
 $d_x$ at low-field phase and $d_z$ at high-field phase. 
 We have mathematically confirmed that the multi-component superconducting phase is absent
because of the idealistic model setting in this paper.  
The conditions for the multi-component superconductivity will be 
discussed elsewhere. In Fig.~S1 in SM, we demonstrate that the reentrant superconductivity happens 
even in usual single-band superconductors with the odd-parity potentials
$\bm{d}_{\bm{k}}=-\bm{d}_{-\bm{k}}$ and $\bm{\alpha}_{\bm{k}}=-\bm{\alpha}_{-\bm{k}}$.
These results indicate that the magnetic configuration in spin space 
$\bm{\alpha} \times \bm{d} \cdot \bm{H} \neq 0$ is essential for the ZFIS and that 
the physical interpretation of the phenomenon using odd-frequency pairs is valid.

We conclude that spin-singlet Cooper pairs 
induced by the interplay between the SOI and a Zeeman field cause 
the reentrant superconductivity.
\textsl{A spin-singlet Cooper pair in high Zeeman fields} is an object 
which does not align with institution.
However, our calculation clearly indicates its existence.
A recent experiment on UTe$_2$ shows a sign of spin-singlet pairs in the high field phase~\cite{rosuel:prx2023}.

\textit{Conclusion}: 
We have studied the stability of a spin-triplet superconducting state described by a $\bm{d}$ vector
in the presence of a spin-orbit interaction $\bm{\alpha}$ and a Zeeman field $\bm{H}$.
The spin-orbit interaction for $\bm{\alpha} \times \bm{d} \neq 0$ 
and Zeeman fields for $\bm{V} \cdot \bm{d} \neq 0$ make the superconducting state 
unstable and decrease the transition temperature. 
The instability is characterized well by the appearance of odd-frequency Cooper pairs 
because they decrease the superfluid density.
We demonstrate that a spin-triplet superconducting state 
exhibits the Zeeman-field-induced superconductivity when 
a scalar chiral product of $ \bm{\alpha} \times \bm{d} \cdot \bm{H}$ remains finite. 
Under the condition, even-frequency spin-singlet even-parity Cooper pairs stabilize 
the spin-triplet superconductivity at high Zeeman fields.
Based on the obtained results, we propose a Hamiltonian of superconducting states which
indicate the reentrant superconductivity in Zeeman fields. 
Although the Jaccarino-Peter compensation has been a reasonable story for the 
Zeeman-field-induced superconductivity, our conclusion provides an 
alternative physical picture for the phenomenon.  
In addition, our theory can be applied also to superconductors preserving 
time-reversal symmetry in the absence of a magnetic field.

\textit{Acknowledgments}: T. S. is supported by JST SPRING, Grant Number JPMJSP2119.
S. I. is supported by a Grant-in-Aid for Early-Career Scientists (JSPS KAKENHI Grant No. JP24K17010).
Y. A. is supported by a Grant-in-Aid for Scientific Research (JSPS KAKENHI Grant No. JP26K0692).

\bibliography{list_2026.bib}

\clearpage
\onecolumngrid
\setcounter{equation}{0}
\setcounter{figure}{0}
\setcounter{table}{0}
\setcounter{page}{1}
\setcounter{section}{0}
\makeatletter
\renewcommand{\theequation}{S\arabic{equation}}
\renewcommand{\thefigure}{S\arabic{figure}}
\renewcommand{\thetable}{S\Roman{table}}
\makeatother

\begin{center}
    \textbf{\large Supplemental Material for ``Reentrant Superconductivity in Zeeman Fields''} \\[.5cm]
    Tomoya Sano, Kota Tabata, Satoshi Ikegaya, and Yasuhiro Asano \\
    \textit{Department of Applied Physics, Hokkaido University, Sapporo 060-8628, Japan}
\end{center}

\vspace{0.5cm}

\section{Solution of Gor'kov equation}\label{sec:solution}
The BdG Hamiltonian is represented by
\begin{align}
H_{\mathrm{BdG}}(\bm{k})  \label{eq:full BdG Hamiltonian}
&=
\begin{bmatrix}
  \xi_{-} + \bm{\alpha} \cdot \hat{\bm{\sigma}} + \bm{V} \cdot \hat{\bm{\sigma}} & 0 & 
  0 &i \bm{d} \cdot \hat{\bm{\sigma}}\, \hat{\sigma}_{y} \\
  0 & \xi_{+} - \bm{\alpha} \cdot \hat{\bm{\sigma}} + \bm{V} \cdot \hat{\bm{\sigma}} &
  -  i \bm{d} \cdot \hat{\bm{\sigma}}\, \hat{\sigma}_{y} & 0 \\
  0 & -i \bm{d} \cdot \hat{\bm{\sigma}}^\ast \, \hat{\sigma}_{y}&
  - \xi_{+} - \bm{\alpha} \cdot \hat{\bm{\sigma}}^{*} - \bm{V} \cdot \hat{\bm{\sigma}}^{*} & 0 \\
  i \bm{d} \cdot \hat{\bm{\sigma}}^\ast\, \hat{\sigma}_{y}\ & 0 &
  0 & - \xi_{-} + \bm{\alpha} \cdot \hat{\bm{\sigma}}^{*} - \bm{V} \cdot \hat{\bm{\sigma}}^{*}
\end{bmatrix} , \\
\xi_{\pm} 
&= 
\frac{1}{2 m}(\bm{k} \pm \bm{K})^2 - \mu, \quad \bm{V} = \mu_{\mathrm{B}} \bm{H},
\end{align}
where $8 \times 8$ matrix structure is derived from spin, valley and particle-hole degree of freedom of
a quasiparticle.
The Hamiltonian can be block-diagonalized into two $4\times 4$ Hamiltonians.
The anomalous Green's function is calculated as a solution of the Gor'kov equation
\begin{align}
\check{G}(\bm{k}, \omega_n) 
=& \check{Z}^{-1}
\left[
\hat{z}_D( i\omega_n - \hat{\xi}_{\bm{k}}) -d^2 ( i\omega_n + \hat{\xi}_{\bm{k}})
- 2( \bm{\alpha}  \cdot \bm{d}, \hat{\rho}_z + \bm{V}  \cdot \bm{d} ) \, \bm{d}  \cdot \hat{\bm{\sigma}} 
\right. \nonumber\\
&+\left. (\hat{z}_D+\bm{d}^2) \bm{\alpha}   \cdot \hat{\bm{\sigma}} \, \hat{\rho}_z
 - (\hat{z}_D- \bm{d}^2) \bm{V}  \cdot \hat{\bm{\sigma}}
 \right],\label{eq:gt0}\\
\check{F}(\bm{k}, \omega_n) 
=& \check{Z}^{-1}
\left[
(\omega_n^2 + \hat{\xi}_{\bm{k}}^2 + \bm{d}^2 - \bm{\alpha}^2 + \bm{V}^2) \, \bm{d} \cdot \hat{\bm{\sigma}}   
 - 2 (\bm{V} \cdot \bm{d}) \bm{V} \cdot \hat{\bm{\sigma}}
 + 2 (\bm{\alpha} \cdot \bm{d}) \bm{\alpha} \cdot \hat{\bm{\sigma}}\right. \nonumber \\
&-2  \hat{\xi}_{\bm{k}}\,  \bm{\alpha} \cdot \bm{d} \hat{\rho}_z  
+ 2 \, i \, \hat{\xi}_{\bm{k}} \, (\bm{V} \times \bm{d}) \cdot \hat{\bm{\sigma}}  \nonumber\\
& \left. +2 i\,  \omega_n\, \bm{V} \cdot \bm{d}  
+ 2 \, \omega_n \, (\bm{\alpha} \times \bm{d}) \cdot \hat{\bm{\sigma}} \, 
\hat{\rho}_z -2 i \, \bm{\alpha} \times \bm{V} \cdot \,
 \bm{d} \, \hat{\rho}_z
 \right]\, \hat{\sigma}_y\, \hat{\rho}_y,\label{eq:ft0}\\
-\undertilde{\check{F}}(\bm{k}, \omega_n) 
=& \hat{\sigma}_y\, \hat{\rho}_y \check{Z}^{-1}
\left[
(\omega_n^2 + \hat{\xi}_{\bm{k}}^2 + \bm{d}^2 - \bm{\alpha}^2 + \bm{V}^2) \, \bm{d} \cdot \hat{\bm{\sigma}}   
 - 2 (\bm{V} \cdot \bm{d}) \bm{V} \cdot \hat{\bm{\sigma}}
 + 2 (\bm{\alpha} \cdot \bm{d}) \bm{\alpha} \cdot \hat{\bm{\sigma}}\right. \nonumber \\
& -2  \hat{\xi}_{\bm{k}}\,  \bm{\alpha} \cdot \bm{d} \hat{\rho}_z  
- 2 \, i \, \hat{\xi}_{\bm{k}} \, (\bm{V} \times \bm{d}) \cdot \hat{\bm{\sigma}}  \nonumber\\
& \left. + 2 i\,  \omega_n\, \bm{V} \cdot \bm{d}  
- 2 \, \omega_n \, (\bm{\alpha} \times \bm{d}) \cdot \hat{\bm{\sigma}} \, 
\hat{\rho}_z 
+2 i \, \bm{\alpha} \times \bm{V} \cdot \,
 \bm{d} \, \hat{\rho}_z
 \right], \label{eq:tilde_ft0}
\end{align} 
\begin{align}
\check{Z}(\bm{k}, \omega_n) = & \left\{ 
\omega_n^2+ \hat{\xi}_{\bm{k}}^2  + \bm{d}^2 - 
\bm{\alpha}^2 + \bm{V}^2\right\}^2 \bm{d}^2 
-4 ( \hat{\xi}_{\bm{k}} \bm{V} -i \omega_n \bm{\alpha} \hat{\rho}_z  )^2 \bm{d}^2 
+ 4 (\bm{\alpha} \cdot \bm{d})^2 (\bm{V}^2 +\bm{d}^2) 
+ 4 (\bm{V} \cdot \bm{d})^2 (\bm{\alpha}^2 -\bm{d}^2) \nonumber\\
& +4 (\bm{\alpha} \times \bm{V} \cdot \bm{d})^2 - 8 (\bm{\alpha} \cdot \bm{d}) (\bm{V} \cdot \bm{d}) 
(\bm{\alpha} \cdot \bm{V}), \label{eq:zt0}\\
=& \hat{\rho}_x\, \check{Z}(-\bm{k}, -\omega_n) \, \hat{\rho}_x,\label{eq:ap_symz}\\
\hat{z}_D=& (i\omega_n + \hat{\xi}_{\bm{k}} )^2 - (\bm{\alpha} \hat{\rho}_z + \bm{V})^2.
\end{align}
The anomalous Green's function is antisymmetric under the operation of interchanging two electrons,
\begin{align}
\check{F}(\bm{k}, \omega_n) = - \hat{\rho}_x \, \check{F}^{\mathrm{T}}(-\bm{k}, -\omega_n) \, \hat{\rho}_x,
\end{align} 
where $\mathrm{T}$ represents the transpose of the Pauli matrices for spin meaning the 
exchange of two spins and $\hat{\rho}_x \cdots \hat{\rho}_x$ represents 
the exchange of the two valleys.
Eq.~\eqref{eq:ap_symz} indicates that $\check{Z}$ is symmetric under such operation.
The three pairing correlations on the first line in Eq.~\eqref{eq:ft0} belong to spin-triplet 
odd-valley-parity symmetry class and are linked to the pair potential 
through the gap equation,
\begin{align}
i \left(\bm{d} \cdot \hat{\bm{\sigma}} \, \hat{\sigma}_y \right)_{\alpha, \beta}
=& - T \sum_{\omega_n} \frac{1}{V_{\mathrm{vol}}} \sum_{\bm{k}}
\sum_{\gamma, \delta} g_{\alpha\, \beta; \gamma\, \delta}
\hat{F}^{(y)}_{\gamma, \delta}(\bm{k}, \omega_n), \label{eq:gapeq}\\
g_{\alpha\, \beta; \gamma\, \delta}=& \sum_{\nu=x, y, z} g_\nu 
\left(i \hat{\sigma}_\nu \, \hat{\sigma}_y \right)_{\alpha, \beta}
\;
\left(i \hat{\sigma}_\nu \, \hat{\sigma}_y \right)^\ast_{\gamma, \delta}, \label{eq:g_conf}
\end{align}
where $\hat{F}^{(y)}$ is the $\hat{\rho}_y$ component of the anomalous Green's function
and $\bm{g}=(g_x, g_y,g_z)$ represents the attractive interaction between two 
electrons at an $s$-wave channel.
To draw Fig.~\ref{fig2}, we assume an attractive interaction at $p$-wave channel 
as usual 
\begin{align}
g_\nu(\bm{k}-\bm{k}^\prime) = g_{\nu} \cos( \theta - \theta^\prime), \; \cos\theta=k_x/k_F, \; 
\sin\theta=k_y/k_F. 
\end{align}
To our knowledge, the two components proportional to $\hat{\xi}_{\bm{k}}$ at the second line in 
Eq.~\eqref{eq:ft0} do not play any important role in stabilizing superconducting states. 
Namely, the presence of these components does not change $T_c$ at all.
The first two components at the last line belong to odd-frequency symmetry class 
and make the superconducting state unstable~\cite{asano:prb2015}.
A Zeeman field parallel to $\bm{d}$ and a SOI perpendicular to $\bm{d}$ 
generate such odd-frequency pairs from the pair potential.
As a result, these components decrease $T_c$.
The last component at the third line in Eq.~\eqref{eq:ft0} belongs to 
even-frequency spin-singlet even-valley-parity symmetry class.
This component stabilizes a superconducting state at high Zeeman fields. 
However, it appears only when the three vectors have a finite spin chiral product 
$\bm{\alpha} \times \bm{H} \cdot \, \bm{d} $.

The anomalous Green's function for $\bm{\alpha} \parallel \bm{H} \parallel \bm{d} $ 
near the transition temperature is represented as
\begin{align}
  \check{F}_\parallel (\bm{k}, \omega_{n})
  =&-\frac{1}{2} \left[
  (\hat{X}_+^{-1} + \hat{X}_-^{-1}) \bm{d}\cdot \hat{\bm{\sigma}} 
  -(\hat{X}_+^{-1} - \hat{X}_-^{-1}) d  \right]
  i\hat{\sigma}_y\, (-i)\hat{\rho}_y, \label{eq:para_f1}\\
  \hat{X}_\pm = & (\omega_n \pm i V)^2 + (\hat{\xi}_{\bm{k}} \mp \alpha\, \hat{\rho}_z) +d^2.
\end{align}
It is possible to analyze the symmetry of Cooper pairs even after summation over $\bm{k}$ 
because all Cooper pairs belong to even-parity $s$-wave symmetry class. 
The first term represents the pairing correlation belonging to 
even-frequency spin-triplet odd-valley parity class and is linked to 
the pair potential through the gap equation.
A Zeeman field induces the pairing correlation belonging to 
odd-frequency spin-singlet odd-valley parity class as shown in the second term.
The anomalous Green's function for $\bm{d} \perp \bm{\alpha}$, $\bm{d} \perp \bm{H}$, and 
$\bm{H} \perp \bm{\alpha}$ near $T_c$ is calculated to be
\begin{align}
  \check{F}_\perp(\bm{k}, \omega_{n})
  =&- \check{Z}_\perp^{-1}
  \left[
(\omega_n^2 + \hat{\xi}_{\bm{k}}^2 - \alpha^2 + V^2) \bm{d}\cdot \hat{\bm{\sigma}}
-2i \hat{\xi}_{\bm{k}} \,  \bm{d} \times \bm{V} \cdot \hat{\bm{\sigma}} 
\right. \nonumber\\
&\left.
+2 \omega_n \bm{\alpha} \times \bm{d} \cdot \hat{\bm{\sigma}} \hat{\rho}_z
+ 2i \bm{\alpha} \times \bm{d} \cdot \bm{V}  \hat{\rho}_z
  \right]
  i\hat{\sigma}_y  (-i)\hat{\rho}_y, \label{eq:perp_f1}\\
  \check{Z}_\perp = & \hat{\xi}_{\bm{k}}^4 + 2 \hat{\xi}_{\bm{k}}^2 (\omega_n^2 - \alpha^2 - 
  V^2)+
 (\omega_n^2 + \alpha^2 + V^2)^2.
\end{align}
The SOI makes the superconducting state unstable because it generates odd-frequency pairing correlation 
at the second line in Eq.~\eqref{eq:perp_f1}.
The last term represents even-frequency spin-singlet even-valley parity Cooper pairs 
which stabilize the superconducting state at high Zeeman fields.

\section{Superfluid weight}\label{sec:weight}

Within the linear response to a static vector potential, the electric current is represents by
$ \bm{j} = - ({n e^2 Q}/{mc})\,  \bm{A}$, 
 where $n$ is the electron density per spin. 
The superfluid weight is defined by~\cite{agd}
\begin{align}
Q =& Q_G + Q_F, \quad
 Q_G = 
 T\sum_{\omega_n}\, \int d\xi_{\bm{k}}\,  \frac{1}{2} \mathrm{Tr}[ \check{G}\, \check{G} - \check{G}_{\mathrm{N}}\, \check{G}_{\mathrm{N}}], \quad 
  Q_F =   
T\sum_{\omega_n}\,  \int d\xi_{\bm{k}} 
\,  \frac{1}{2} \mathrm{Tr}[- \check{F}\, \undertilde{\check{F}} ], \label{eq:qf_def}
\end{align}
with $\check{G}_{\mathrm{N}}$ being the Green's function in the normal state.
Since we find $Q_G=Q_F$ in this paper, we discuss how each component in the 
anomalous Green's function contributes to $Q_F$.
 Substituting the electric current into the Maxwell equation 
 $ \nabla \times \bm{H} = \frac{4\pi}{c} \bm{j}$, 
 the Meissner screening length $\lambda$ increases with decreasing $Q_F$ as $\lambda \propto Q_F^{-1/2}$.
A superconductor indicates the diamagnetic response to magnetic fields as long as $Q_F>0$.
Thus, the superfluid weight also represents the stability of the superconducting states.
We will show that the even-frequency pairing correlations increase 
$Q_F$, whereas the odd-frequency components decrease $Q_F$.
Thus, it is often said that even-frequency Cooper pairs (odd-frequency Cooper pairs) 
indicate diamagnetic (paramagnetic) response to magnetic fields.
The coefficient $b$ in Eq.~(7) is proportional to $Q_F$ with a same sign~\cite{sato:prb2024}.
Therefore, the transition to the superconducting state becomes a first-order for $Q_F<0$.

The superfluid weight $Q_F$ is calculated from 
the product of the Green's function in Eqs.~\eqref{eq:ft0} and \eqref{eq:tilde_ft0}.
Almost all the cross terms vanish due to the summation over the Matsubara frequency, 
summation over $\bm{k}$, and the trace over spin plus valley spaces. 
The first three terms in Eq.~\eqref{eq:ft0} are the principal pairing correlation 
and couple only to the first three terms in Eq.~\eqref{eq:tilde_ft0}.
These terms belonging to even-frequency symmetry class are linked to the pair potential.
The remaining five terms in Eq.~\eqref{eq:ft0} couple only to their 
particle-hole conjugate in Eq.~\eqref{eq:tilde_ft0}.
It is easy to confirm that the even-frequency components increases $Q_F$ and 
odd-frequency component decrease $Q_F$.
As a consequence, the appearance of odd-frequency pairs makes the superconducting states 
unstable and decreases $T_c$~\cite{asano:prb2015,sato:prb2024}.

For the parallel configuration, the superfluid weight results in
\begin{align}
  q(T, V)  &=  q_{d}(T, V) + q_{\mathrm{odd}}(T, V), \\
  q_{d}
  &=
  2 \pi T \sum_{\omega_{n} > 0}
  \frac{2 \omega^{4}_{n} - \omega^{2}_{n} V^{2} + V^{4}}
  {2 \omega_{n} (\omega^{2}_{n} + V^{2})^{3}} , \quad
  q_{\mathrm{odd}}
  =
  - 2 \pi T \sum_{\omega_{n} > 0}
  \frac{5 \omega^{2}_{n} V^{2} + V^{4}}
  {2 \omega_{n} (\omega^{2}_{n} + V^{2})^{3}}. \label{eq:qeo_para}
\end{align}
The superfluid weight of the principal pairing correlation $q_d$ is 
positive, whereas that 
of induced odd-frequency pairing correlations $q_{\mathrm{odd}}$ is negative.
 Fig.~2(a) in the text shows $q_{d}(T, H_c)$ 
 and $q_{\mathrm{odd}}(T, H_c)$ as a function of temperatures 
 along the phase boundary displayed in Fig.~1(a), where 
$H_c$ is obtained from the data points in Fig.~1(a).
The vertical axis is normalized to 
\begin{align}
q_{\mathrm{BCS}}(T) =  2\pi T \sum_{\omega_n>0} \omega_n^{-3},
\end{align}
at $T=T_0$.
The results in Fig.~2(a) show the total superfluid is negative for $T<0.556T_0$. 
As a consequence, the transition to superconducting phase becomes a first-order 
at such low temperatures in Fig.~1(a).
The results in Eq.~\eqref{eq:qeo_para} are exactly equal to the superfluid weights 
of a spin-singlet $s$-wave superconductor in a Zeeman field.

For the perpendicular configuration,
the superfluid weights are calculated from the anomalous Green's function in Eq.~\eqref{eq:perp_f1}
\begin{align}
q(T,V)=&  
q_d(T,V)+ q_{\mathrm{odd}}(T,V) + q_{\perp}(T,V)
,\\
q_d= & 2 \pi T\sum_{\omega_n>0}  \frac{2\omega_n^6 -( \bm{\alpha}^2- 6  \bm{V}^2) \omega_n^4 
+( \bm{\alpha}^4 + 2 \bm{\alpha}^2  \bm{V}^2 +6 \bm{V}^4 ) \omega_n^2 
+  \bm{V}^2 (\bm{\alpha}^2 +\bm{V}^2 )
(\bm{\alpha}^2 +2 \bm{V}^2 )}{2  \omega_n^{3} (\omega_n^2 + \bm{\alpha}^2 + \bm{V}^2)^3},\\
q_{\mathrm{odd}} = & - 2 \pi T\sum_{\omega_n>0}  \frac{\omega_n^2 \bm{\alpha}^2( 5 \omega_n^2 +  \bm{\alpha}^2 + \bm{V}^2 )}
{ 2  \omega_n^{3} (\omega_n^2 + \bm{\alpha}^2 + \bm{V}^2)^3 },\quad
q_{\perp} =  2 \pi T\sum_{\omega_n>0} \frac{ \bm{\alpha}^2  \bm{V}^2 ( 5 \omega_n^2 + \bm{\alpha}^2 + \bm{V}^2 ) }
{2 \omega_n^{3} (\omega_n^2 + \bm{\alpha}^2 + \bm{V}^2)^3}. \label{eq:induce_q_perp}
\end{align}
The first line in Eq.~\eqref{eq:perp_f1} gives the weight of spin-triplet pairs $q_d$. 
The superfluid weight due to odd-frequency
component is negative as shown in $q_{\mathrm{odd}}$. 
The spin-singlet pairing correlation at the last term in Eq.~\eqref{eq:perp_f1} increases the 
weight as shown in $q_\perp$. 
We fix a temperature at $T=0.5T_0$ in Fig.~1(b) for $\alpha = 1.5T_0$ 
and plot these superfluid weights as a function of the Zeeman potential in Fig.~2(b).
The results show that 
$|q_{\mathrm{odd}}|$ decreases and $q_\perp$ increases with increasing $H$. 
They almost cancel to each other around the critical Zeeman potential 
to ZFIS is $\mu_{\mathrm{B}} H_c = 1.65 T_0$ as indicated by a dotted line. 
As a result of such compensation, the superfluid weight of the principal component 
$q_d$ increases with $H$. 
The monotonic increase of the total superfluid weight implies 
a Zeeman field stabilizes the superconducting state in the perpendicular configuration.
Indeed, the transition temperature increases with increasing the Zeeman potential.
We find $q>0$ for all the phase boundary in Fig.~1(b). 
Thus, the transition to the superconducting phase is a second-order~\cite{sato:prb2024}.

\section{Reentrant superconductivity in single-band superconductors}

\begin{figure}[thb]
  \centering
  \includegraphics[width = 8cm]{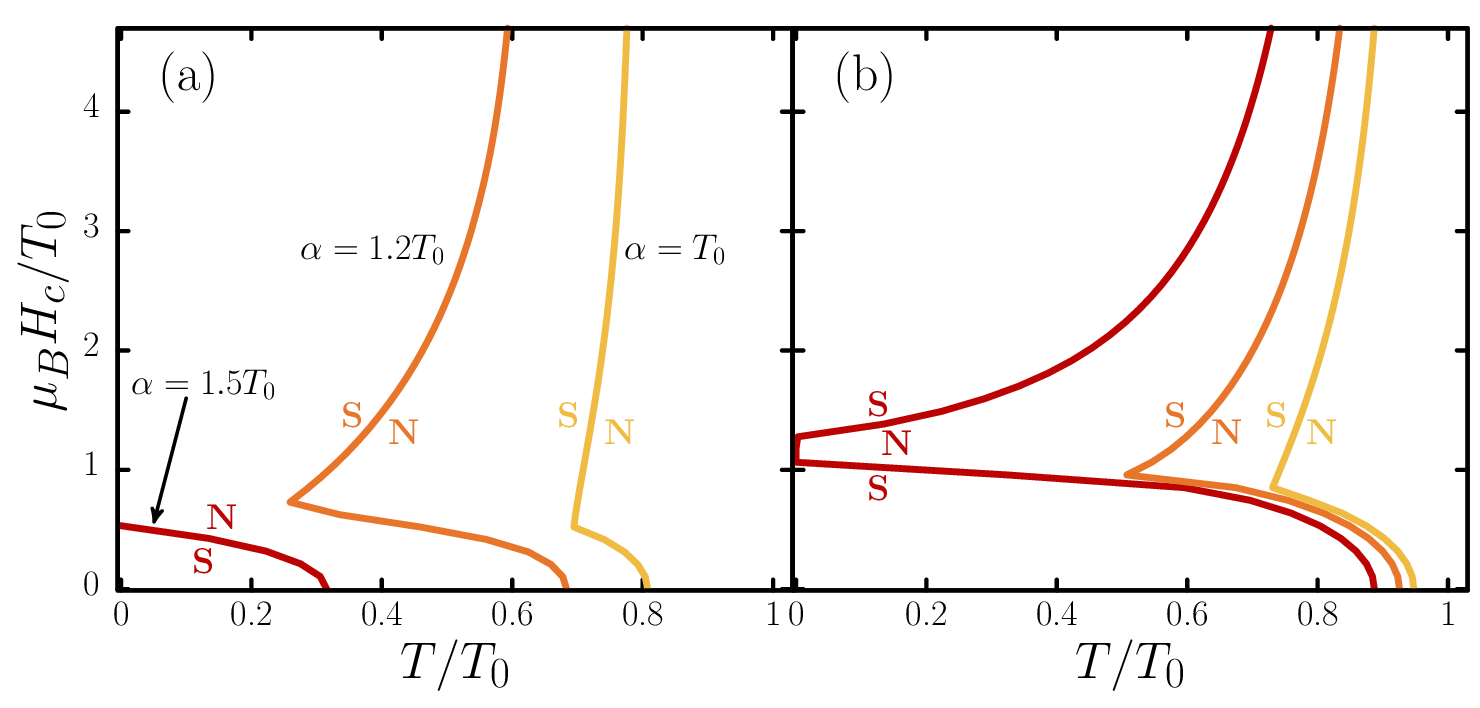}
  \caption{
    The critical magnetic field $H_{c}$ is plotted as a function of temperature 
	for single-band spin-triplet superconductors with Rashba SOI.
The pair potential is chosen as Eq.~\eqref{eq:d2a} in (a) and Eq.~\eqref{eq:d2b} in (b).
The low-field (high-filed) phase is the single component spin-triplet superconductivity with 
$d_x \neq 0$ ($d_z \neq 0$).
  } \label{fig2}
\end{figure}

In the text, we discuss the mechanism of the reentrant superconductivity in Zeeman field by 
using the analytical expressions of the superfluid weights obtained in the two-valley model.
Here, we demonstrate the reentrant superconductivity in a usual single-band SC in two-dimension,
\begin{align}
  \check{H}_{\mathrm{N}} (\bm{k}) 
  =&
  \left(\frac{\bm{k}^2}{2 m}   - \mu\right) \hat{\sigma}_{0}
  +
  \bm{\alpha}_{\bm{k}} \cdot \hat{\bm{\sigma}} 
  +
  V \hat{\sigma}_x, \label{eq:hn2}\\
\hat{\Delta}(\bm{k})=& i \bm{d}_{\bm{k}} \cdot \hat{\bm{\sigma}}\, i \hat{\sigma}_y,\quad  
\bm{\alpha}_{\bm{k}} = \alpha\left( \bar{k}_y \bm{e}_x -  \bar{k}_x \bm{e}_y\right), 
  \end{align}
where we assume the Rashba SOI and $\bar{k}_j=k_j / k_F$ for $j = x, y$.
We consider the two pair potentials as
\begin{subequations}
\begin{align}
\bm{d}_{\bm{k}} =& d_x \bar{k}_x \bm{e}_x + d_z \bar{k}_y \bm{e}_z, \label{eq:d2a}\\
\bm{d}_{\bm{k}} =& d_x \bar{k}_y \bm{e}_x + d_z \bar{k}_x \bm{e}_z. \label{eq:d2b}\
\end{align}
  \end{subequations}
We assume that the attractive interaction at $p$-wave channel in the gap equation.  
The phase diagrams for Eqs.~\eqref{eq:d2a} and \eqref{eq:d2b} are shown in 
Fig.~\ref{fig2}(a) and (b), respectively.
The results for $\alpha=1.5T_0$ in (a) show usual single superconducting phase, 
whereas those in (b) show two separated superconducting phases.
The phase diagram depends sensitively on the relative $\bm{k}$ dependence 
between $\bm{\alpha}_{\bm{k}}$ and $\bm{d}_{\bm{k}}$ for large $\alpha$.
Such tendency seems to be weaker for smaller $\alpha$.
 The reentrant superconductivity
can be seen in the results for $\alpha=1.2T_0$ and $\alpha=T_0$ in both Figs.~\ref{fig2}(a) and (b). 

\end{document}